\documentclass[
showpacs,
floatfix,
aps,
pre,
amsmath,
twocolumn,
groupaddress,
]{revtex4-1}
\usepackage{amsmath,amssymb}
\usepackage{amscd, latexsym}
\usepackage{mathrsfs}
\usepackage{graphicx}
\usepackage{epstopdf}
\usepackage{cancel}
\usepackage{amsfonts}
\usepackage{exscale}
\usepackage{dcolumn}
\usepackage{bm}
\usepackage{color}
\usepackage{xcolor}
 \usepackage{natbib}
\usepackage{lmodern}
\usepackage[T1]{fontenc}
\usepackage{verbatim}
\usepackage{float}
\usepackage {hyperref}
\hypersetup{colorlinks=true}

\usepackage[normalem]{ulem}

\newcommand{\ket}[1]{\left| #1 \right>} 
\newcommand{\bra}[1]{\left\langle #1 \right|} 

\newcommand{\tr}[1]{{\rm Tr}\left\{ #1 \right\}}

\makeatletter
\newsavebox{\@brx}
\newcommand{\llangle}[1][]{\savebox{\@brx}{\(\m@th{#1\langle}\)}%
  \mathopen{\copy\@brx\kern-0.5\wd\@brx\usebox{\@brx}}}
\newcommand{\rrangle}[1][]{\savebox{\@brx}{\(\m@th{#1\rangle}\)}%
  \mathclose{\copy\@brx\kern-0.5\wd\@brx\usebox{\@brx}}}
\makeatother

\renewcommand{\vec}[1]{\bm{#1}}

\begin{document}

\title{Response of a quantum disordered spin system to a local periodic drive}

\author{A. Bar\i\c{s} \"Ozg\"uler}
\affiliation{Department of Physics, University of Wisconsin--Madison, Madison, WI 53706}
\author{Canran Xu}
\affiliation{Department of Physics, University of Wisconsin--Madison, Madison, WI 53706}
\author{Maxim G. Vavilov}
\affiliation{Department of Physics, University of Wisconsin--Madison, Madison, WI 53706}

\date{October 2, 2019}

\begin{abstract} 
We consider a one-dimensional spin chain system with quenched disorder and in the presence of a local periodic drive. We study the time evolution of the system in the Floquet basis and evaluate the fidelity susceptibility, which is a measure of how a given state changes under a small perturbation, of states to a weak periodic drive. We demonstrate  that the statistical properties of the fidelity susceptibility over different disorder realizations can be used to identify two phases of the system:  (1) the many-body localized phase, in which the susceptibility exhibits long tails while its average value decreases rapidly as disorder increases; and (2) the ergodic phase, in which the susceptibility distribution is narrow and its average value weakly depends on disorder. This distinction in the average value of the susceptibility between the two phases develops readily for systems with ten or more spins.  Therefore, recently built networks of  qubits subject to a local drive can simulate  dynamics of a system in the many-body localization regime. We also show that the spin accumulation speed is correlated with the fidelity susceptibility and can also be used to distinguish  the two phases. 
\end{abstract}


\maketitle
\section{Introduction}\label{sec1}

Studies of electron localization in disordered systems have over half century history starting with the seminal work by Anderson \cite{Anderson1958}. A system of non-interacting electrons in one and two dimensions exhibit localization at zero temperature as follows from the scaling considerations \cite{Abrahams1979}.
The role of electron-electron interactions, however, is ambiguous.  The onset of localization, known as weak localization \cite{Altshuler1980}, is destroyed by electron--electron interaction at finite temperatures \cite{Altshuler1982}
as the interaction results in dephasing of electron wave functions.  At the same time, electron--electron interactions give rise to the Coulomb gap at the Fermi energy, driving the system to localization \cite{Efros1975}. Theory of many-body localization (MBL) in disordered many-body system of interacting electrons was put forward in the work of Basko, Aleiner and Altshuler \cite{Basko2006}.  This paper proposed an infinite order perturbation theory in the electron--electron interaction  and determined an energy threshold for localization.  Below the threshold, the interactions between electrons cannot facilitate electron hopping between localized single electron states and systems remain localized.  As energy of the electron system increases above the threshold value, a large phase space of the system allows electrons to rearrange and form an extended many-electron quantum state.  This many-electron quantum state corresponds to dephasing in a single electron language.

Further focus of MBL studies was to understand interacting many-body spin systems with random field. Interacting electrons and spin-1/2 chains are closely related models. The spinless electron system  can be mapped onto XXZ chain via Jordan-Wigner transformation \cite{coleman2015introduction}. The onsite energy in the fermionic system corresponds to random $z$-field in the spin chain model. 

Both fermionic systems and spin chains with disorder have been  shown to exhibit MBL transition in Refs \cite{Pal2010, Monthus2010, deLuca2013, Schiulaz2014, Lev2014, Oganesyan2014, Abanin2017, Alet2018}. This transition between localized and ergodic regimes can be characterized via entropy growth \cite{Bardarson2012,Serbyn2013}, localization length \cite{Serbyn2013}, energy spectrum \cite{Oganesyan2007, xu2017response}, local integrals of motion \cite{Serbyn2013b, Kim2014,ponte2015, Chandran2015, Ros2015420} and entanglement \cite{Serbyn2013, Bardarson2012, Nanduri2014, Serbyn2016}. In the ergodic phase, the level statistics obeys a level spacing similar to the Wigner-Dyson distribution with level repulsion. The dynamic susceptibility is large. In the localized phase, on the other hand, the Hamiltonian of the system shows localized behavior, as the level spacing is characterized by a Poisson distribution with high probability to find two levels with a small level separation, and the dynamic susceptibility vanishes \cite{Atas2013a, xu2017response}. If a system is prepared in a product state, the entanglement entropy for its subsystems gets saturated quickly for ergodic regimes, and the saturation value is given by the Page value, which is proportional to system size $L$ \cite{PageValue_PhysRevLett.71.1291, PageValue_PhysRevB.95.094206}. However, in the MBL regime, the entanglement entropy gets saturated in exponentially long time \cite{Bardarson2012,Serbyn2013} but the saturation value still scales linearly with the system size.

\begin{figure}[!htbp]
\begin{centering}
\includegraphics[width = \columnwidth]{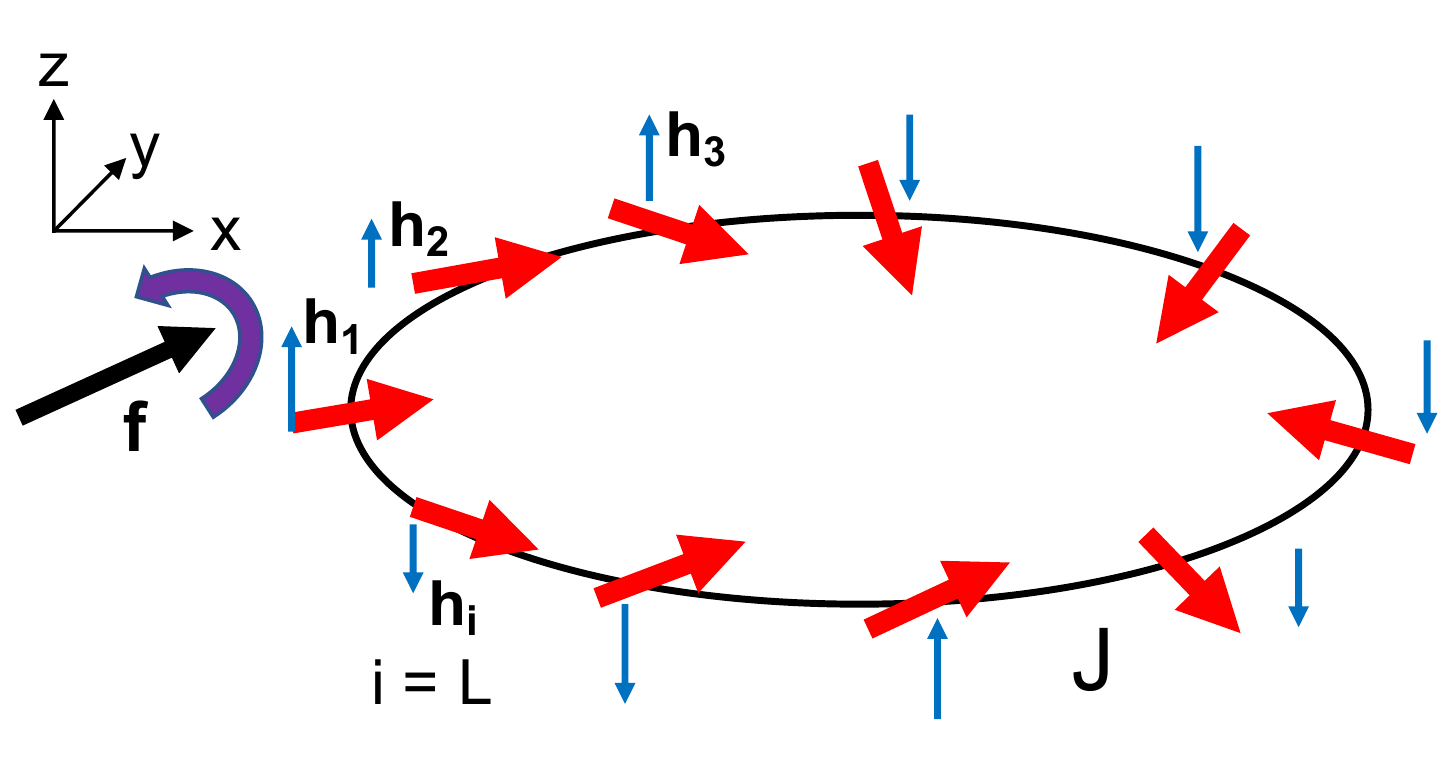}
\par\end{centering}
\caption{\label{fig:SystemSketch}
(Color online) 
Heisenberg spin-1/2 chain system with quenched disorder $\{h_l\}$ in $z$-direction. $\{h_l\}$ is defined by the uniform distribution within the interval $|h_l|\leq W$. $W$ is the disorder strength and the interaction strength between the nearest neighbors are given by the unitless parameter $J = 1$. There is a local ac drive with strength $f$ on the spin labeled by $i = 1$ in the $x$-$y$ plane rotating with drive frequency $\omega$ in the anticlockwise direction. 
}
\end{figure}

The purpose of this paper is to evaluate an experimentally accessible method to observe MBL phases by using a local harmonic drive on one of the spin with period $\tau=2\pi/\omega$. We present our results for the short time scales when the system may not have reached its saturation value yet. We consider a one-dimensional Heisenberg spin chain system with quenched disorder driven by a local ac field. The static Heisenberg Hamiltonian with the periodic boundary condition $\vec{\sigma}^{(L+1)}=\vec{\sigma}^{(1)}$ is given by 
\begin{equation}
H_{0}=
\sum_{l=1}^{L}\left[ J \bm{\sigma}^{(l)} \bm{\sigma}^{(l+1)} +
h_{l}\sigma^{(l)}_{z}
\right].
\label{eq:static_h}
\end{equation}
Here, $\bm{\sigma}^{(l)}$ is the vector of Pauli matrices for spin at site $l$. The onsite fields $h_l$ are independent random fields, uniformly distributed in the range $[-W,W]$, where $W$ is the disorder strength of the system.  At weak disorder, $W\lesssim 3J$, the system is in the ergodic regime and has several characteristics reminiscent of conduction phase of a disordered metal. According to previous numerical studies, the transition from the ergodic regime to the localized regime takes place at $W=W_c\simeq 3J$ \cite{Pal2010, Luitz2015}.  We use $J$ as a fundamental unit and set $J = 1$ throughout the rest of this paper.

The system with Hamiltonian \eqref{eq:static_h} conserves the total $z$-component of spin 
\begin{equation}
S_z = \frac{1}{2}\sum_l\sigma_z^{(l)},
\label{eq:Sz_total}
\end{equation}
A transverse ac drive is applied to a single spin,
\begin{equation}
V(t)= f[\cos(\omega t)\sigma_{x}^{(1)}+\sin(\omega t)\sigma_{y}^{(1)}],\label{eq:drive}
\end{equation}
which breaks the conservation of $S_z$.  Here, $f$ denotes the strength of the drive, $\omega$ is the drive frequency and $\tau = 2\pi/\omega$ is the period of the drive.  

We investigate time evolution of the system described by the time-dependent full Hamiltonian
\begin{equation}
H(t)=H_{0}+V(t).
\label{eq:dynamic_h}
\end{equation}
We apply Floquet theory to analyze the system's response to the periodic drive. Periodic time-dependent Hamiltonians were also studied in \cite{d2013many,lazarides2015fate, Ponte2014,ponte2015,abanin2016theory,rehn2016periodic,Zhang2016a} using Floquet analysis.
We use exact numerical diagonalization and the time dynamics of the Hamiltonian for a system size $L\lesssim 18$.  Further increase of the system size requires significant increase in computing power and memory requirements. 
We perform analysis of fidelity susceptibility \cite{GU2010} and change in system dynamics of total spin as the strength of disorder changes from weak to strong. 
An experimental platform to study these quantities could be one of the available quantum hardware for quantum computing, such as optical lattices \cite{schreiber2015observation} trapped ions \cite{Smith2016}, Rydberg atoms \cite{Bernien2017}, ultracold atoms \cite{Bordia2017}, gmon system \cite{Neill2018} and fluxonium qubits \cite{nguyen2018highcoherence}.

Fidelity susceptibility was previously used to study phase transition  \cite{de2019ground, Li2017, Hu2016, Hu2017, Monthus2017}. In this paper, we study fidelity susceptibility as a measure of overlap between the two quantum states $|\langle \psi_{f=0}|\psi_{f\neq0} \rangle|^2$ that evolve with or without drive from the same initial state $|\psi_i\rangle$, where $\overline{(...)}$ stands for the  average over initial states $\ket{\psi_i}$. For weak drive, the quantum displacement is proportional to the fidelity susceptibility. Evolution of an initial state may follow different paths in the Hilbert space depending on the phase of many-body systems.  An important factor that defines the quantum displacement between the two final states is disorder. When the disorder is weak, the distance between the two final states is large. However, for strong disorder, localization occurs and the distance vanishes.

The local drive \eqref{eq:drive} breaks $S_z$-conservation law. We show that spin accumulation in response to the drive could be a viable  experimental method to distinguish between localized and ergodic regimes. The variance of operator $S_z$ with respect to an arbitrary quantum state $\ket{\psi(t)}$ of the system at time $t$ is
\begin{equation}
\delta S_z^2(t)= \langle S_z^2(t) \rangle - \langle S_z (t)\rangle^2,
\label{eq:Variance_Sz}
\end{equation}
where, $\langle A (t)\rangle$ is defined as $\langle A (t)\rangle \equiv \langle \psi(t) | A | \psi(t) \rangle$. We  perform analysis of statistical properties of the spin accumulation $\delta S_z^2(t)$ over disorder realizations. We study average of $\delta S_z^2(t)$ as a function of time $t = n\tau$, where $n$ is the number of periods. The statistics of spin accumulation is significantly different for the ergodic and MBL regimes and the difference between the spin accumulation over time can be used to distinguish between the two regimes. The change in $\delta S_z^2(t)$ after one period can be identified as the total spin diffusion coefficient. 
We compare the quantum displacement at one period with the diffusion coefficient  $\delta S_z^2(\tau)$ and show that they have similar behavior.
We analyze the distribution of the diffusion coefficient for different disorder strengths. The distributions are different for the MBL and ergodic regimes. The diffusion coefficient is large and the distribution is narrow for weak disorder, whereas the diffusion coefficient is small and the distribution is wide and has a long tail for the strong disorder.

The paper is organized as follows.  
In Sec. \ref{Floquet_Rep}, we first introduce the model and derive a formalism that can be used to analyze the ac driving scenario for the model within Floquet theory. In Sec. \ref{FidSus_WeakDrive}, we study the response of the quantum system by evaluating the  quantum displacement between the evolution of the system with and without an ac drive. Then in Sec. \ref{Stats_Fid_Sus}, we provide the numerical results for the statistical properties of the fidelity susceptibility and compare with our analytical estimations. Finally, in Sec. \ref{Spin_TimeEvolution}, we study time dynamics and statistical properties of total spin in $z$-direction. That provides an experimentally feasible way to distinguish between ergodic and MBL phases.

\section{Evolution in the Floquet Representation}\label{Floquet_Rep}

For a periodic drive, the evolution operator $U(t=n\tau)$ after $n$ periods can be represented as the $n$th power of the Floquet operator $U_f(\tau)$ per one period $\tau = 2\pi/\omega$:  $U(t=n\tau)=U_f^n$.  The Floquet operator is unitary and has a set of eigenvectors, that form a Floquet basis:
\begin{equation}
U_f \ket{\alpha}= e^{-i\Omega_\alpha t}\ket{\alpha}, \label{eq:floquet_basis}
\end{equation}
where we use Greek indices to denote the Floquet basis, $\alpha=1,\dots,2^L$, and $\Omega_\alpha$ are quasienergies.  After $n$ periods of the drive, the system evolves  from its initial state $\ket{\psi_0}$ to the state
\begin{equation}
\ket{\psi(n\tau)}=U_f^n \ket{\psi_0}, \quad
U_f=e^{-i\Omega_\alpha t}\ket{\alpha}\bra{\alpha}\label{eq:floquet_evolution}
\end{equation}
with $U_f$ written in the Floquet basis. 

To evaluate the Floquet operator, we notice that the transformation 
\begin{equation}
U_1(t)=\exp\left(\frac{i\omega t}{2} \sum_{l=1}^L \sigma_z^{(l)}\right)\label{eq:U1}
\end{equation}
removes explicit time-dependence in the full Hamiltonian of the system, Eq.~\eqref{eq:dynamic_h}:
\begin{equation}
\tilde{H}=U_1HU_1^{\dagger}-iU_1\dot
U_1^{\dagger}=\tilde H_{0}+ f \sigma_{x}^{(1)},\quad \tilde H_{0}=H_0-\omega \, S_z.\label{eq:transform}
\end{equation}
After this transformation, the Floquet operator can be defined as an exponent of time-independent Hermitian operator 
\begin{equation}
U_f = \exp(-i \tilde{H} \tau) \, U_1(\tau) = (-1)^L \, \exp(-i \tilde{H}  \tau),
\label{eq:U_full}
\end{equation}
We notice that for $f=0$, the Floquet states $\ket{\alpha}$ and eigenvectors of stationary Hamiltonian $\tilde H_0$ as well as quasienergies $\Omega_\alpha$ and energies $\tilde E_i$ coincide, 
\begin{equation}
\langle i\ket{\alpha_{f=0}}=\delta_{i\alpha},\quad
\Omega_\alpha=\tilde E_\alpha ({\rm mod}\  2\pi/\tau).
\label{eq:quasi_spectrum}
\end{equation} 

Using Eq.~\eqref{eq:U_full}, we find that the Floquet basis is simply given by the eigenstates of the transformed Hamiltonian ~\eqref{eq:transform} \cite{NoteFloquet}. Response of quantum disordered spin systems to a local periodic drive \cite{Ponte2014} and global drive \cite{d2013many,lazarides2015fate,ponte2015,abanin2016theory,rehn2016periodic,Zhang2016a} were also studied, where Hamiltonian is switched between two different operators periodically in time. Differently, we consider local ac drive in this paper.

The effect of a harmonic drive on a state of the system can be defined by the displacement of this state $\ket{\psi_f(\tau)}$ after one period of the drive from the free evolution over the period $\tau$ of the same  state $\ket{\psi_0(\tau)}$. For an arbitrary initial state $\ket{\psi_i}$, the state after one period is  $\ket{\psi_f(\tau)}=U_f\ket{\psi_i}$ for a harmonic drive with amplitude $f$, and 
$\ket{\psi_0(\tau)}=U_0\ket{\psi_i}$, where $U_0=U_{f\to 0}=\exp(-i H_0\tau)$. The corresponding overlap between the two states is measured by the real part of the Fubini-Study metric and is simply determined by the overlap of these two states: 
\begin{equation}
\begin{aligned}
F_{\psi_i} =\left|
\langle\psi_0(\tau)\ket{\psi_f(\tau)}
\right|^2 = \left|
\langle\psi_i| {\cal U}  \ket{\psi_i}
\right|^2,
\label{eq:fidelity}
\end{aligned}
\end{equation}
where we introduced a unitary operator 
\begin{equation}
{\cal U} = U_0^\dagger U_f\label{eq:Udef}
\end{equation} 
representing a mismatch between the evolution of the system with and without drive. Fubini-Study metric is known as quantum geometric tensor in the adiabatic limit. The imaginary part of the quantum geometric tensor gives the Berry curvature. Both real and imaginary parts of the quantum geometric tensor can be used as susceptibility to measure phase transitions  \cite{gritsev2012dynamical}. Here, to identify phases, we use fidelity susceptibility for the weak drive, which will be defined in the next section.

We characterize a typical response of an arbitrary state to the drive over a single period in terms of the overlap $F_{\psi_i}$. The quantum fidelity is given as $F = \overline{F_{\psi_i}}$, where $\overline{(...)}$ stands for the  average over initial states $\ket{\psi_i}$. The corresponding average, known as a quantum fidelity between two unitary operations, is defined in terms of operator ${\cal U}$ as \cite{pedersen2007fidelity}
\begin{equation}
F= \displaystyle \frac{M+\left|tr({\cal U})\right|^2}{M(M+1)}\label{eq:QuantumFidelity},
\end{equation} 
where $M=2^L$ is the dimensionality of the Hilbert space. We define quantum displacement between the two final states after one period as in the following:
\begin{equation}
\varepsilon \equiv 1-F.
\label{eq:QuantumDisplacement}
\end{equation} 
In the weak drive limit, this quantity is proportional to the fidelity susceptibility and its analysis is given in the next section.

We calculate the matrix element of ${\cal U}$ taken between the energy eigenstate $\ket{i}$ of the static Hamiltonian $H_0$ and the Floquet state $\ket{\alpha}$, which has the form
\begin{equation}
\bra{i}{\cal U}\ket{\alpha}=A_i^\alpha\exp(-i(\Omega_\alpha-E_i)\tau), \quad A_i^\alpha= \langle i\ket{\alpha},
\label{eq:Uai}
\end{equation}
where $A_i^\alpha$ is the overlap amplitudes between energy eigenstates of the static Hamiltonian and the Floquet states.
This relation leads to the matrix elements of ${\cal U}$ in the energy eigenstate basis of $H_0$:
\begin{equation}
{\cal U}_{ij}=\bra{i}{\cal U}\ket{j}=\sum_\alpha  \exp(-i(\Omega_\alpha-E_i)\tau) 
A_i^\alpha (A_j^\alpha)^*.
\label{eq:Uij}
\end{equation}
According to this equation, the evolution of the system reduces to a search of the components $A_i^{\alpha}$ of the Floquet states in the basis of the static Hamiltonian, and the corresponding eigenenergies and quasienergies. Below we present numerical evaluation of these matrix elements and argue that the statistical properties $A_i^{\alpha}$ change across the crossover from the ergodic to MBL regimes.

For quantitative analysis of the effect of the drive on the system, we consider a Hermitian matrix 
\begin{equation}
{\cal T}=i\frac{1-{\cal U}}{1+{\cal U}}\label{eq:T}
\end{equation}
instead of the unitary matrix ${\cal U}$.  A simple choice of the norm as $\propto tr({\cal T}^2)$ can be interpreted as the power of the drive applied to the system.  This is especially meaningful in the limit of weak drive when ${\cal T}$ is linear in the drive amplitude $f$. 
In its eigenvector basis, operator ${\cal U}$ is presented by a diagonal matrix with elements $e^{i\delta_a}$ ($a=1,\dots, M$) and ${\cal T}$ is also diagonal with diagonal elements $[{\cal T}]_{aa}=\tan(\delta/2)$.  The norm of ${\cal T}$ is
\begin{equation}
tr({\cal T}^2)=\sum_{a=1}^M\tan^2\frac{\delta_a}{2}\label{eq:normT^2}
\end{equation}
and $tr({\cal T}^2)\to\infty$ when one of the scattering phases reaches the unitary limit, $\delta_a=\pi$, 
 so that corresponding eigenvector $\ket{\alpha}$ of ${\cal U}$ completely flips just after a single period of the drive, ${\cal U}\ket{\alpha}=-\ket{\alpha}$.  This strong effect of the system states does not necessarily reduce fidelity ${F}$ Eq.~\eqref{eq:QuantumFidelity}. However, the system rearrangement over energy states $\ket{i}$ of the stationary Hamiltonian $H_0$ per cycle of the drive becomes significant if $\langle i\ket{\alpha}\neq 0$ for many states $\ket{i}$.

Utilizing Eq.~\eqref{eq:Uai} and ~\eqref{eq:T}, we can write the system of linear equations for the Floquet amplitudes $A_i^{\alpha}$:
\begin{equation}
\sum_j
\bra{i}\frac{\tan((\Omega_\alpha-E_i)\tau/2)+{\cal T}}{1-i{\cal T}}\ket{j}A_{j}^{\alpha}=0.
\label{eq:hoppingH}
\end{equation}
This equation can be reduced to a hopping problem \cite{Ponte2014} of a particle with on-site energy 
$\tan((\Omega_\alpha-E_i)\tau/2)$ and hopping amplitude ${\cal T}$ between sites in the Hilbert space:
\begin{equation}
\left[
\tan\frac{(\Omega_\alpha-E_i)\tau}{2}+{\cal T}
\right]\ket{\chi_\alpha}=0,\label{eq:T_equation2}
\end{equation}
where $\ket{\chi_\alpha}=\sum_j (1-i{\cal T})^{-1}\ket{j}A_j^{\alpha}$ is an eigenstate at zero energy existing for a set of quasienergies $\Omega_\alpha$ of the Floquet operator $U_f$.  Equation~\eqref{eq:hoppingH} is in particular useful in the limit of weak drive when it establishes a simple relation between the Floquet amplitudes $A_i^{\alpha}$ and hopping amplitudes ${\cal T}_{ij}$, which is derived in the next section.

\section{Fidelity susceptibility at weak drive}
\label{FidSus_WeakDrive}

Two initial same states are evolved under unperturbed and perturbed Hamiltonians for a period.   We calculate the quantum displacement $\varepsilon$ given by Eq.\eqref{eq:QuantumDisplacement} between the two final states after a period, which is independent of the given initial state and depends only on the mismatch between the energy eigenstates of the unperturbed Hamiltonian and Floquet basis. 
When the drive strength $f$ is small, we can write the Maclaurin series expansion for the fidelity in  Eq.~\eqref{eq:fidelity} around $f = 0$: 
\begin{equation}
F = 1 - \frac{f^2}{2} \chi_F + ...,
\label{eq:Fidelity_Expansion}
\end{equation}
and neglect the higher order terms. Here $\chi_F$ is defined as \textit{fidelity susceptibility} and it is the second derivative of the fidelity with respect to the drive amplitude $f$ \cite{GU2010}. In the small $f$ limit, $\chi_F$ can be written in terms of fidelity $F$:
\begin{equation}
\chi_F = 2 \, (1- F)/f^2 =  2 \, \varepsilon/f^2 .
\label{eq:Fidelity_Susceptibility}
\end{equation}
Note that $\chi_F$ is proportional to the quantum displacement $\varepsilon$ given by Eq.~\eqref{eq:QuantumDisplacement}.

In this section, we consider in detail the limit of weak external drive and take into account only terms that are linear in drive amplitude $f$ in the hopping matrix ${\cal T}$ and the unitary matrix ${\cal U}$. 
First, we expand  the operator ${\cal U}$, defined by Eq.~\eqref{eq:Udef}, to the lowest order in $f$, and obtain the following expression for the hopping matrix:
\begin{equation}
\begin{split}
& {\cal T} =  -\frac{if\tau}{2}  \\
&\times
\left(\sigma_x^{(1)}+i\tau[\tilde H_0,\sigma_x^{(1)}] +\frac{(i\tau)^2 }{2!}
[\tilde H_0,[\tilde  H_0,\sigma_x^{(1)}]]+\dots
\right).
\end{split}\label{eq:T_expansion}
\end{equation}
This expression indicates that the matrix elements of ${\cal T}_{ij}$ can be easily written in the eigenstate basis of Hamiltonian $\tilde H_0$ in terms of $\bra{i}\sigma_x^{(1)}\ket{j}$.  Here we present an alternative derivation of ${\cal T}_{ij}$.  We consider Eq.~\eqref{eq:hoppingH}  up to the first order in ${\cal T}$ and apply Eq.~\eqref{eq:quasi_spectrum} to find a relation between off diagonal elements of matrices $A_i^{j\neq i}$ and ${\cal T}_{ij}$ written in the eigenstate basis:
\begin{equation}
{\cal T}_{ij}=iA_{i}^{\alpha\to j}\sin\frac{\pi(\tilde E_i-\tilde E_j)}{\omega}e^{i \pi(\tilde E_i-\tilde E_j)/\omega}.\label{eq:TinA}
\end{equation}
To the lowest order in $f$, overlap between Floquet states and eigenstates of $\tilde H_0$ can be evaluated from the first order perturbation theory as $A_{i\neq j}^{\alpha\rightarrow j}= f \bra{i}\sigma_x^{(1)}\ket{j}/(\tilde E_i-\tilde E_j)$. Note that while the difference between eigenenergies $E_i$ of $H_0$ and $\tilde E_i$ of $\tilde H_0$ are not important in Eq.~\eqref{eq:TinA}, this difference is important in the denominator of 
$A_{i\neq j}^{\alpha\rightarrow j}$, which represents transition between states with different values of total spin along the $z$-axis, due to absorption or emission of energy $\hbar\omega$. We obtain the following expression for matrix elements of the hopping matrix in the basis of eigenstates of $H_0$ that coincides with eigenstates of $\tilde H_0$:
\begin{equation}
{\cal T}_{ij}=f\frac{\bra{i}\sigma_x^{(1)}\ket{j}}{\tilde E_i-\tilde E_j}\sin\frac{\pi(\tilde E_i-\tilde E_j)}{\omega}e^{i \pi(\tilde E_i-\tilde E_j)/\omega}.
\label{eq:Tij}
\end{equation}

At weak drive, $\mathcal{U}=1+2i{\cal T}-2{\cal T}^{2}+\dots$ and we obtain an expression for the average fidelity 
\begin{equation}
F  =\frac{M+M^2-4M \, tr({\cal T}^2)}{M(M+1)}.
\label{eq:F_p}
\end{equation} 
The quantum displacement can be regarded as the average displacement of the states per period of the drive. In the expression above, we disregarded terms that contain $(\tr{{\cal T}})^2 $ since $\tr{{\cal T}}$ vanishes.

We apply  Eq.~\eqref{eq:TinA} to argue that the quantum displacement, $\varepsilon$, is a universal, $M-$independent measure of the effect of a harmonic drive on the system in either ergodic or MBL regimes. We write
\begin{equation}
\varepsilon \leq \frac{1}{M}\sum_i\sum_{\alpha\neq i} \left|A_i^\alpha\right|^2 =\frac{\sum_iP^{(i)}_{\rm esc}}{M}=\overline{P^{(i)}_{\rm esc}},
\label{eq:p_ave}
\end{equation}
where $\overline{P^{(i)}_{\rm esc}}$ is the escape probability $P^{(i)}_{\rm esc}=1-|A_i^i|^2$ of the system from initial state $\ket{i}$ at long drive time, averaged over states $\ket{i}$.

We can provide a more accurate estimate of quantum displacement by applying Eq.~\eqref{eq:Tij}:
\begin{equation}
\varepsilon \simeq \frac{\pi^2 f^2}{\omega^2 M}\sum_{i\neq j}
\frac{\sin^2(\pi(\tilde E_i- \tilde E_j)/\omega)}{[\pi(\tilde E_i-\tilde E_j)/\omega]^2}
\left|\bra{i}\sigma_x^{(1)}\ket{j}
\right|^2.
\label{eq:pweak}
\end{equation}
First, we evaluate the average value of quantum displacement over realizations of the random magnetic field for ergodic regime of weak disorder $W\lesssim 3J$.  At frequencies of the drive exceeding the mean level spacing
we omit the energy dependent factor. Also, a typical matrix element for $i\neq j$ can be estimated as $\sum_{i\neq j} \left|\bra{i}\sigma_x^{(1)}\ket{j} \right|^2 \simeq M$. So, after these approximations, we can write the quantum displacement as:
\begin{equation}
\langle\langle \varepsilon \rangle\rangle \, \propto \frac{\pi^2 f^2}{\omega^2},
\label{eq:barescale}
\end{equation}
where $\langle\langle...\rangle\rangle$ represents the disorder average throughout the paper.

In the limit of strong disorder, the distribution of quantum displacement is more complicated. As we demonstrate below from numerical analysis, the distribution becomes extremely wide and its average value actually loses its meaning.  More meaningful is the distribution of the logarithm of quantum displacement, $\lg(\varepsilon)$, and its disorder average $\langle\langle\lg(\varepsilon)\rangle\rangle$. The exponential of $\langle\langle\lg(\varepsilon)\rangle\rangle$ gives the typical value (geometric mean) for the quantum displacement. Note that $\lg$ shows $log_{10}$ throughout the text. The logarithmic distribution is a common characteristic of strongly disordered, glassy systems that exhibit a wide hierarchy of scales \cite{spinglassreview_RevModPhys.58.801}. In our case, the broad distribution is formed due to rival realizations of the 
random magnetic field.  For some realizations, the spin states are strongly localized and effectively decoupled from the rest of the system, for other realizations the system develops a resonance between spins in the chain and may result in the quantum displacement exceeding the average displacement in the ergodic regime, \textit{cf.} Eq.~\eqref{eq:barescale}. 

The contribution from configurations representing localized spins dominates for average value of $\lg(\varepsilon)$, and results in monotonically decreasing value of   $\langle\langle\lg(\varepsilon)\rangle\rangle$.  For localized states case when the local magnetic field for a driven spin is strong, $|h_1|\gg J$, the eigenstates $\ket{i}$ are factorized and we can reduce the evaluation of quantum displacement in Eq.~\eqref{eq:pweak} as
\begin{equation}
\varepsilon \simeq \frac{f^2}{M}\frac{M}{2}
\frac{\sin^2(\pi h_1/\omega)}{h_1^2}
\left|\bra{\downarrow}\sigma_x^{(1)}\ket{\uparrow}
\right|^2.
\label{eq:ploc}
\end{equation}
Assuming that the localized configurations give the main contribution to $\langle\langle \lg(\varepsilon) \rangle\rangle$, we integrate $\lg(\varepsilon)$ given by Eq.~\eqref{eq:ploc} over uniformly distributed $h_l$ and obtain
\begin{equation}
\langle\langle \lg(\varepsilon) \rangle\rangle \, \propto 2\lg\frac{J}{W}.
\label{eq:lgpstrong}
\end{equation}
We note that our estimates for $\varepsilon$ in the  limit of weak or strong disorder are independent of the dimensionality of the Hilbert space $M=2^L$, see Eqs.~\eqref{eq:barescale} and \eqref{eq:lgpstrong}.

\section{Statistical properties of the fidelity susceptibility}\label{Stats_Fid_Sus}

In this section, we numerically evaluate the quantum displacement $\varepsilon$, which determines the fidelity susceptibility in  Eq.~\eqref{eq:Fidelity_Susceptibility}. We choose the drive frequency $\omega = J$ for the simulations. We do not expect the results to be very different for $\omega$ comparable to $J$. As pointed in Sec. \ref{conclusions} Discussion and Conclusions;  new phases of matter can arise for large $\omega$, and the limit of small $\omega$ can be studied as a dc perturbation to the Heisenberg Hamiltonian \cite{xu2017response}. We calculate the fidelity directly from Eq.~\eqref{eq:QuantumFidelity}, by computing the matrix exponents for evolution matrices $U_0$ and $U_f$, and therefore, our computation is not restricted to the weak drive limit considered in the previous section. For $f\ll J$, we obtained the bilinear response of $\varepsilon\propto f^2$ and recover all relations between the Floquet amplitudes $A^\alpha_i$, quasienergies and matrix elements of $\sigma_x^{(1)}$ between unperturbed eigenstates of $H_0$ that we discussed in the previous section.  We also observed that the bilinear regime is satisfied for average value of $\varepsilon$ or $\lg(\varepsilon)$ for $f\lesssim J$, and chose $f=J/\sqrt{10}$ for analysis of $\varepsilon$ at different values of disorder strength $W$.
This choice of $f$ allows us to compare some conclusions from the previous section with the numerical results, and at the same time demonstrates that the properties of $\varepsilon$ remain similar at moderate drive amplitudes, $f\simeq J$.  At stronger drive, multiphoton processes become important and their analysis deserve a separate discussion.

\begin{figure}[!htbp]
\begin{centering}
\includegraphics[width = \columnwidth]{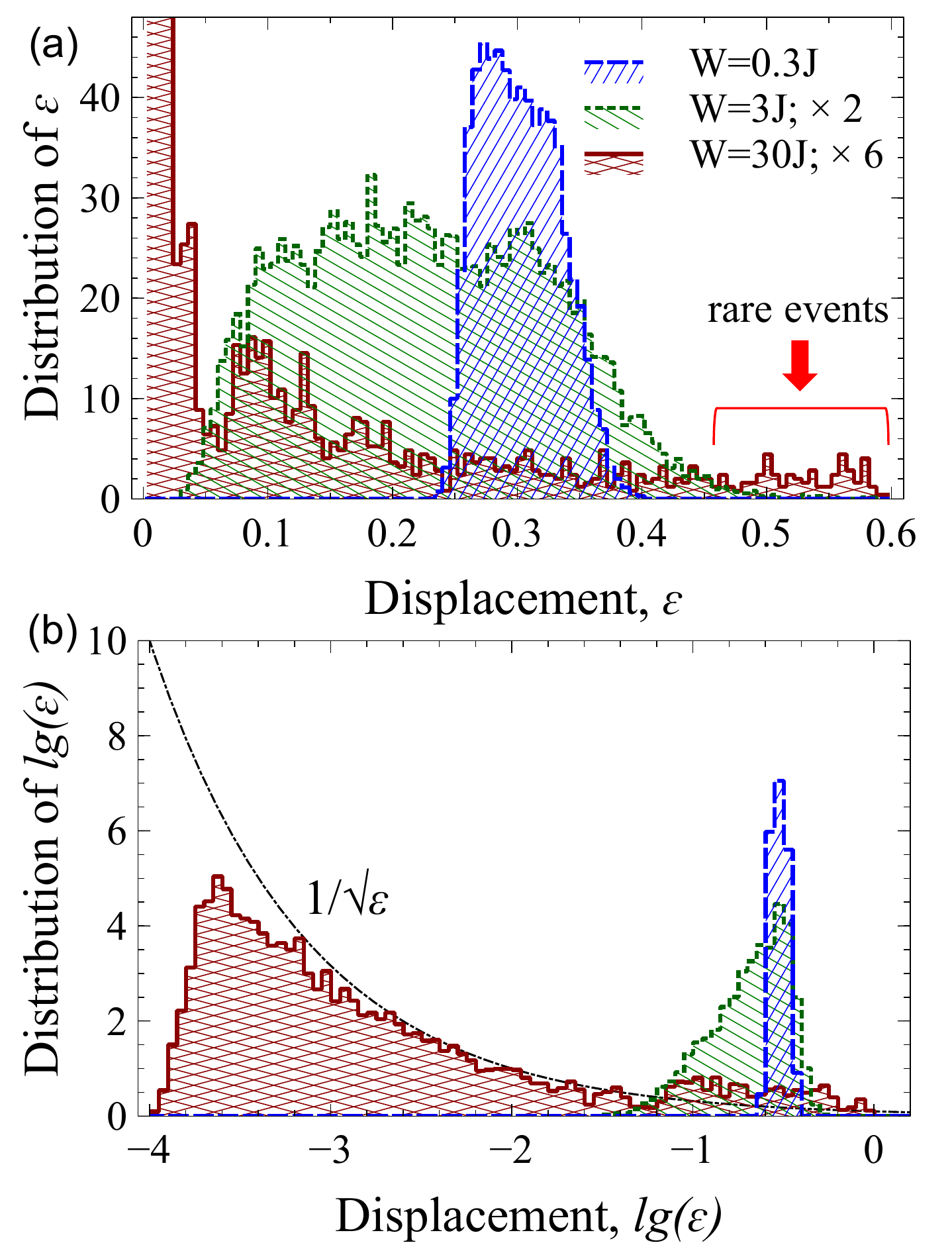}
\par\end{centering}
\caption{\label{fig:hist}
(Color online) 
(a) Distribution of quantum displacement $\varepsilon$ over $N=10^4$ realizations of the random magnetic field $h_l$ for a system with $L=12$ spins. The top panel shows the distribution of the displacement itself for $W/J=0.3$ (blue long-dashed line), $W/J=3$ (green short-dashed line), and $W/J=30$ (red solid line). Distributions for strong disorder have exponentially large tails. Rare events appear for the strong disorder.  (b)  Logarithm of the distribution of $\varepsilon$ for the same three values of disorder as in (a). The dash-dotted line represents the slope $\sim 10^{-1/2(\lg (\varepsilon))} = 1\sqrt{\epsilon}$. The drive amplitude $f=J/\sqrt{10}$ and $\omega=J$. $\lg$ shows $log_{10}$ throughout the text.  We scaled the distribution curves for $W/J=3$ by factor two and for $W/J=30$ by factor six.}
\end{figure}

First, we study the probability distribution $P(\varepsilon)$
at the drive strength $f = J/\sqrt{10}$ over ensemble realizations of the random fields $\{h_l\}$ defined by a uniform distribution within the interval $|h_l|\leq W$.  Because our numerical analysis required evolution of matrix exponent and inverting matrices, to reach a large number of realizations $N=10^4$, we took the system size $L=12$.  We present the normalized histograms in Fig.~\ref{fig:hist} for weak, moderate and strong strengths of disorder.  As the strength of disorder increases, the distribution broadens and the peak shifts to smaller values of $\varepsilon$. However, while more realizations have smaller values of $\varepsilon$, there are some realizations at moderate disorder that exhibit $\varepsilon$ exceeding maximal values of $\varepsilon$ in weakly disordered system, see the tail to the right in Fig.~\ref{fig:hist}(a).  This behavior becomes even more pronounced at strong disorder, $W=30J$, when the distribution covers extremely small values of $\varepsilon$, but its tail extends to larger values of $\varepsilon$ than the values found for weak and moderate disorder, see Fig.~\ref{fig:hist}(b).

We characterize the distribution in the strong disorder limit by $\lg(\varepsilon)$.  In such logarithmic presentation, it is possible to fit all distributions of three cases of weak, moderate and strong disorder on the same plot, as shown in Fig.~\ref{fig:hist}(b).  At strong disorder, distribution of $\lg(\varepsilon)$ shows that in most realizations, the quantum displacement is significantly reduced below its values for the ergodic regime. At the same time, we find the tail that extends to larger values of $\varepsilon$, which does not happen at weaker disorder.  In these rare events, quantum displacement $\varepsilon$ takes values closer to 1 and our bilinear analysis is not applicable, in particular, relation \eqref{eq:F_p} is no longer valid.  For realizations with large values of $\varepsilon$, the system exhibits occasional resonances between spins in the chain that lead to strong coupling of the drive to the spin system.  In this case, the spin system subject to a drive strongly deviates  from its free evolution. 

We plot the distribution of the logarithm of the quantum displacement in the limit of strong disorder in Fig.~\ref{fig:hist}(b) and observe that the right slope is consistent with 1/$\sqrt{\epsilon}$.  This behavior implies that the probability distribution function for $\varepsilon$ decays as a power law $\propto (\varepsilon)^{-3/2}$, and we conclude that the distribution of quantum displacement is Pareto type. Such slow power law decrease makes the cumulants ill-defined, including the expectation value, unless the power law has upper cutoff.  According to Fig.~\ref{fig:hist}(b), the power law terminates at sufficiently large displacement, making the expectation value of displacement over disorder sensitive to the rare large realizations of disorder.  This sensitivity to rare fluctuations of displacement does not allow us to numerically study average value of displacement at strong disorder, as even for a very large number of samples, $N\gtrsim 10^4$ for smaller systems, $L=6$, the average value of displacement does not converge well.  

\begin{figure}[!htbp]
\begin{centering}
\includegraphics[width=\columnwidth]{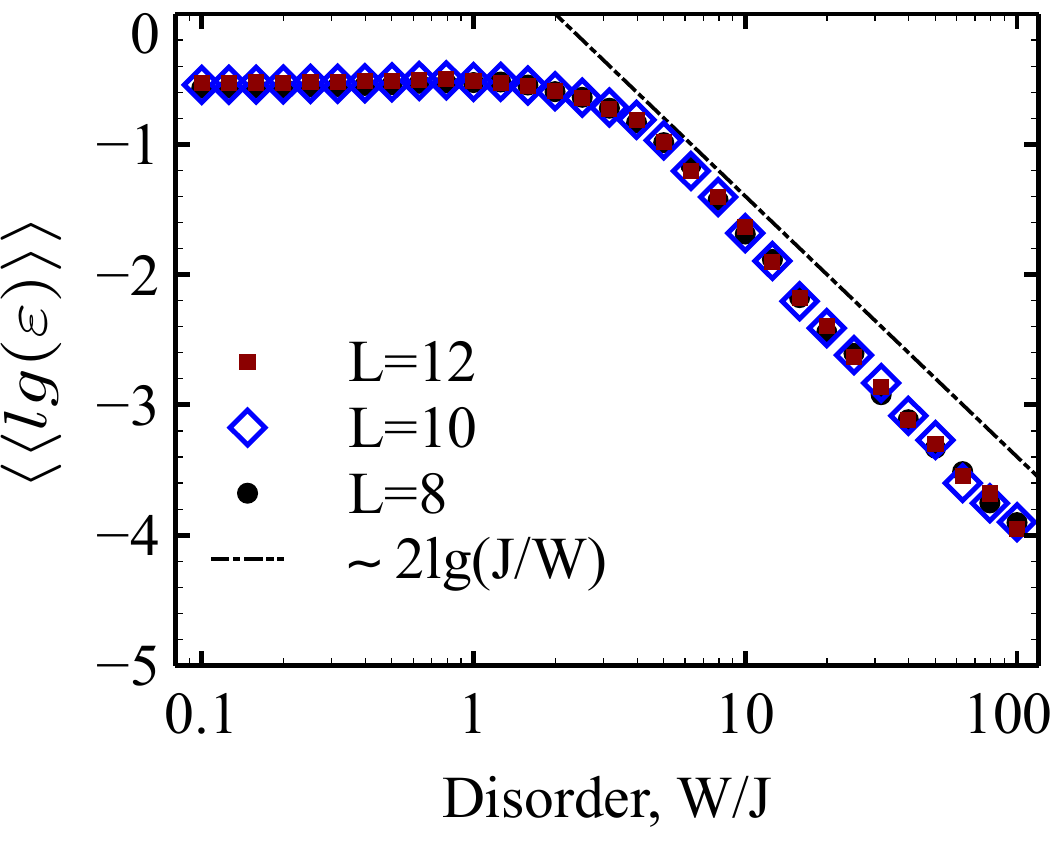}
\par\end{centering}
\caption{\label{fig:avlogT2}
(Color online)
Average of the logarithm of quantum displacement, $\lg (\varepsilon)$, as a function of disorder strength $W$ for a spin system of size  $L=8$ (circles), $L=10$ (squares) and $L=12$ (diamonds).
The average is evaluated over $N=10^3$ disorder samples for $L=8,\ 10,\ 12$. The drive amplitude $f=J/\sqrt{10}$ and $\omega=J$. $\langle\langle\rangle\rangle$ shows the disorder average throughout the paper.
}
\end{figure}

To characterize the effect of disorder strength on the quantum displacement, we numerically evaluate the disorder average of $\lg(\varepsilon)$, which is shown as $\langle\langle \lg(\varepsilon) \rangle\rangle$.  The result is presented in Fig.~\ref{fig:avlogT2}.  We observe that $\langle\langle {\lg(\varepsilon)} \rangle\rangle$ does not strongly depend on the system size $L$, as points for $L=8,\ 10$ and $12$ are aligned along the same curve.  At weak disorder, $\langle\langle {\lg(\varepsilon)} \rangle\rangle$ changes weakly with disorder strength, as demonstrated by different values of $\langle\langle {\lg(\varepsilon)} \rangle\rangle$ at the plateaus for disorder strength corresponding to the ergodic regime with $W\lesssim 3J$.  At stronger disorder, in the localization regime $W\gtrsim 3J$, $\langle\langle {\lg(\varepsilon)} \rangle\rangle$ decreases linearly as $\sim 2\lg(J/W)$, in agreement with estimate \eqref{eq:lgpstrong}.

\section{Time evolution of the total spin}
\label{Spin_TimeEvolution}

In this section, we describe a  technique to distinguish between ergodic and MBL phases using the total spin projection in $z$-direction $S_z$, given by Eq.\eqref{eq:Sz_total}. It has been shown that magnetization can be a probe to distinguish between ergodic and MBL phases \cite{Vasseur2014}. Here, we study the variance of total spin in $z$-direction that gives the measure of localization for a given state \cite{QuantumVariance_PhysRevB.94.075121}. The total spin projection in $z$-direction is a conserved quantum number of $H_0$, Eq.~\eqref{eq:static_h}. When there is a local periodic drive perpendicular to $z$-direction, $S_z$ is not conserved anymore. The value of $S_z$ with respect to time depends on the strength of the random field $W$. For the variance of $S_z$ given by Eq.~\eqref{eq:Variance_Sz}, $\delta S_z^2(t)$, we observe different statistics for the ergodic and MBL phases.

We choose the initial state as a product state with $S_z = 0$. Such product states can be shown as $| \psi \rangle = | \{ \sigma_i \} \rangle$ with $\sigma_i = \pm 1$, $\sum_i \sigma_i = 0$, where +1 represents spin up and -1 represents spin down for even system size $L$. There are $L !/ ((L/2)!)^2$ product states with $S_z = 0$. For systems of size up to $L = 12$, it is computationally feasible to take average $\overline{\delta S_z^2(t)}$ (product state average is shown by overline) over all product states along with disorder average. For the sizes beyond $L = 12$, we took  average over some group of  randomly selected product states. Even a small group of samples can be useful to identify the phase of the system. By analyzing statistical dynamics of product states, we can study the ergodic and MBL phases. By using time dynamics, one can simulate larger systems comparing to the spectral analysis because exact diagonalization is computationally more intensive.

\begin{figure}[!htbp]
\begin{centering}
\includegraphics[width=\columnwidth]{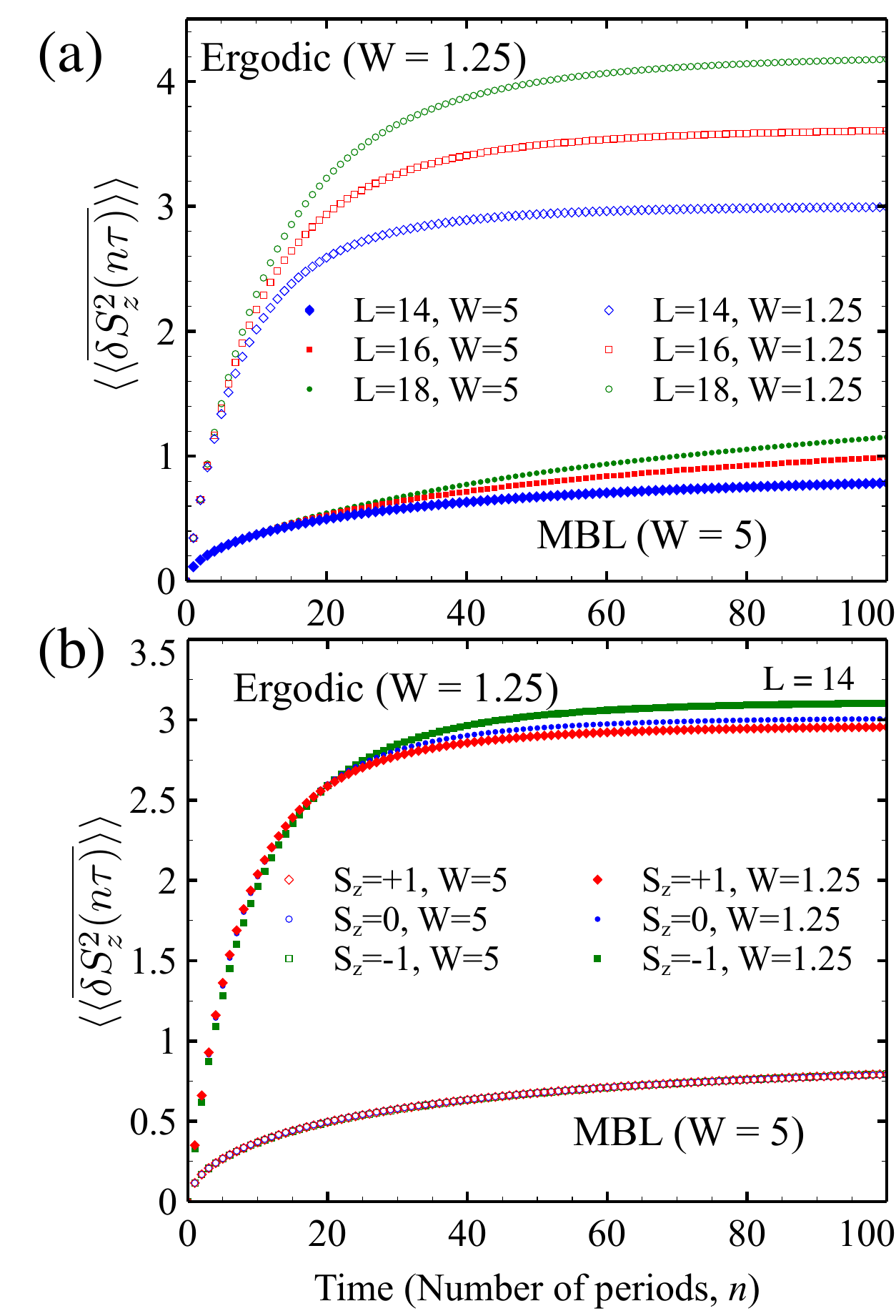}
\par\end{centering}
\caption{\label{fig:VarSz_vs_time}
(Color online)
(a) Average $\delta S_z^2(t)$ as a function of time for a spin system of size $L = 14$ (diamonds), $L = 16$ (squares) and $L = 18$ (circles). Curves for $W = 1.25$ have filled and for $W = 5$ have unfilled markers. The averages are performed over $10^3$ realizations of disorder for all system sizes and $10^3$ product states for $L = 14$, 150 product states for $L = 16$ and 60 product states for $L = 18$. The overline shows the product state average throughout the paper. (b) Average $\delta S_z^2(t)$ as a function of time for a spin system of size $L = 14$ and $W$ = 1.25 or 5. Results are compared for the initial product state with $S_z = \pm 1$ and 0 at $t = 0$. 100 product states and $10^3$ disorder averages are considered for all cases. 
}
\end{figure}

Short time growth of $\delta S_z^2(t)$ can identify the phase of the system \cite{NoteOTOC}. Fig.~\ref{fig:VarSz_vs_time} shows how the average variance $\langle\langle \overline{\delta S_z^2(n\tau)} \rangle\rangle$ changes with respect to the number of periods, $n$ \cite{Note_Integrability}. 
Average is taken over product states (shown by overline) and disorder (shown by double angle brackets). In the ergodic regime, the variance changes quickly for the initial periods and reaches a saturation point for longer times. For $L = 14$, the saturation point is reached in less than one hundred cycles of drive. For larger systems, it takes more time to reach the saturation point. 
One can estimate based on the decreasing rate of change of the variance with time that it does not take exponentially long time to reach the saturation for systems with $L = 16$ and 18 in the ergodic regime. However, in the MBL regime, the variance increases slowly and based on the monotonous increase rate one can estimate that it takes much more time to reach a saturation point comparing to the ergodic case. In addition, the variance change in the MBL regime is less sensitive to the system size than in ergodic regime. In Fig.~\ref{fig:VarSz_vs_time}(b), we demonstrated for different initial conditions and product states ($S_z$ = 0 vs. $\pm$ 1) that one can still distinguish between ergodic and MBL regimes regardless of the initial $S_z$ choices. In MBL regime ($W$ = 5), the spin accumulation takes almost the same values and the curves are aligned with each other. In ergodic regime ($W$ = 1.25), the spin accumulation for the three different initial $S_z$ values slightly differ. The reason of this slight difference between $S_z = \pm$ 1 is the sine term in Eq.\eqref{eq:drive}, which is an odd function and breaks the symmetry with respect to the local field rotation direction.

\begin{figure}[!htbp]
\begin{centering}
\includegraphics[width=0.97\columnwidth]{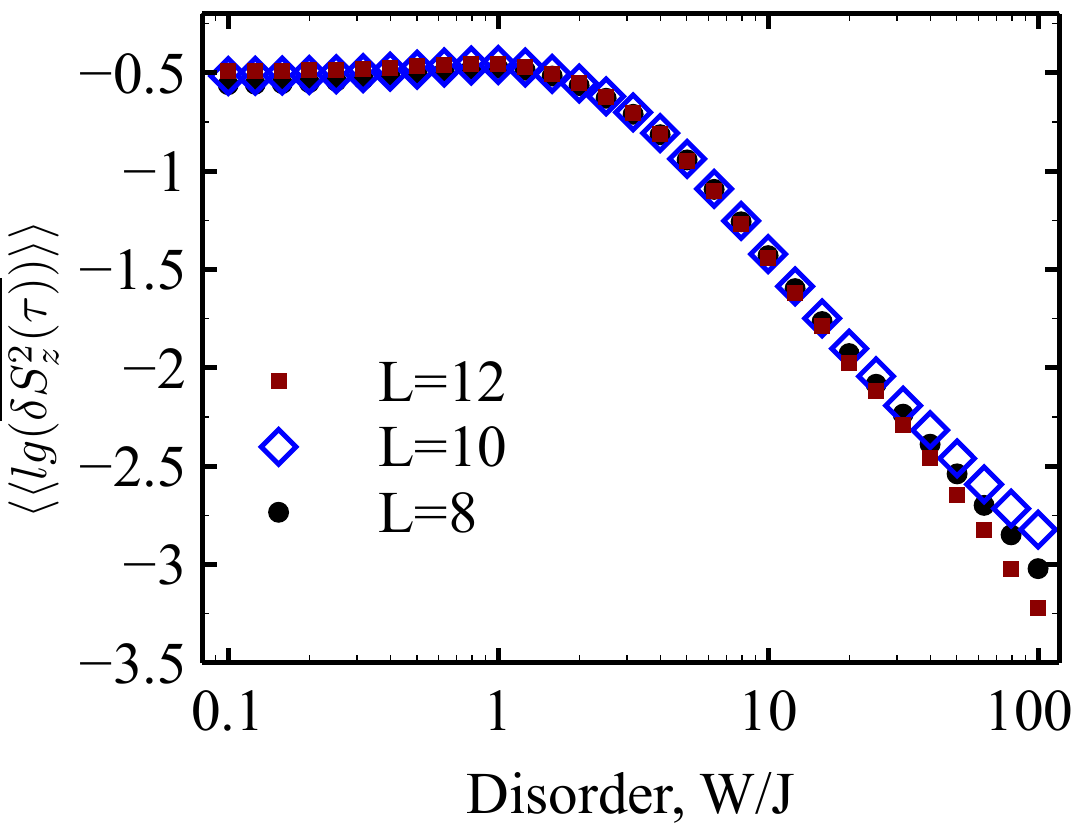}
\par\end{centering}
\caption{\label{fig:VarSz_vs_W}
(Color online)
Average $\delta S_z^2(\tau)$ as a function of $\lg(W/J)$ for a spin system of size  $L=8$ (circles), $L=10$ (squares) and $L=12$ (diamonds). Time = 1 period. $f=J/\sqrt{10}$ and $\omega=J$. The averages are performed over $10^4$ disorder realizations for $L = 8$, $10^3$ disorder realizations for $L = 10$, $L = 12$. All product states are considered for all system sizes for product state averaging.  
}
\end{figure}

In Fig.~\ref{fig:VarSz_vs_W}, we show how the average of logarithm of the variance, $\langle\langle \lg(\overline{ \delta S_z^2(\tau)}) \rangle\rangle$, changes with respect to the disorder strength $W$. Time is fixed at one period, $\tau$. The variance curves in Fig.~\ref{fig:VarSz_vs_W} shows similar properties as the quantum displacement curves in Fig.~\ref{fig:avlogT2}. $\delta S_z^2(t)$ changes weakly with disorder strength at weak disorder ($W \lesssim 3J$), whereas it decreases linearly with $lg( W/J)$ at stronger disorder ($W \gtrsim 3J$). Similar to the quantum displacement, $\delta S_z^2(t)$ also does not strongly depend on the system size $L$.

\begin{figure}[!htbp]
\begin{centering}
\includegraphics[width=0.97\columnwidth]{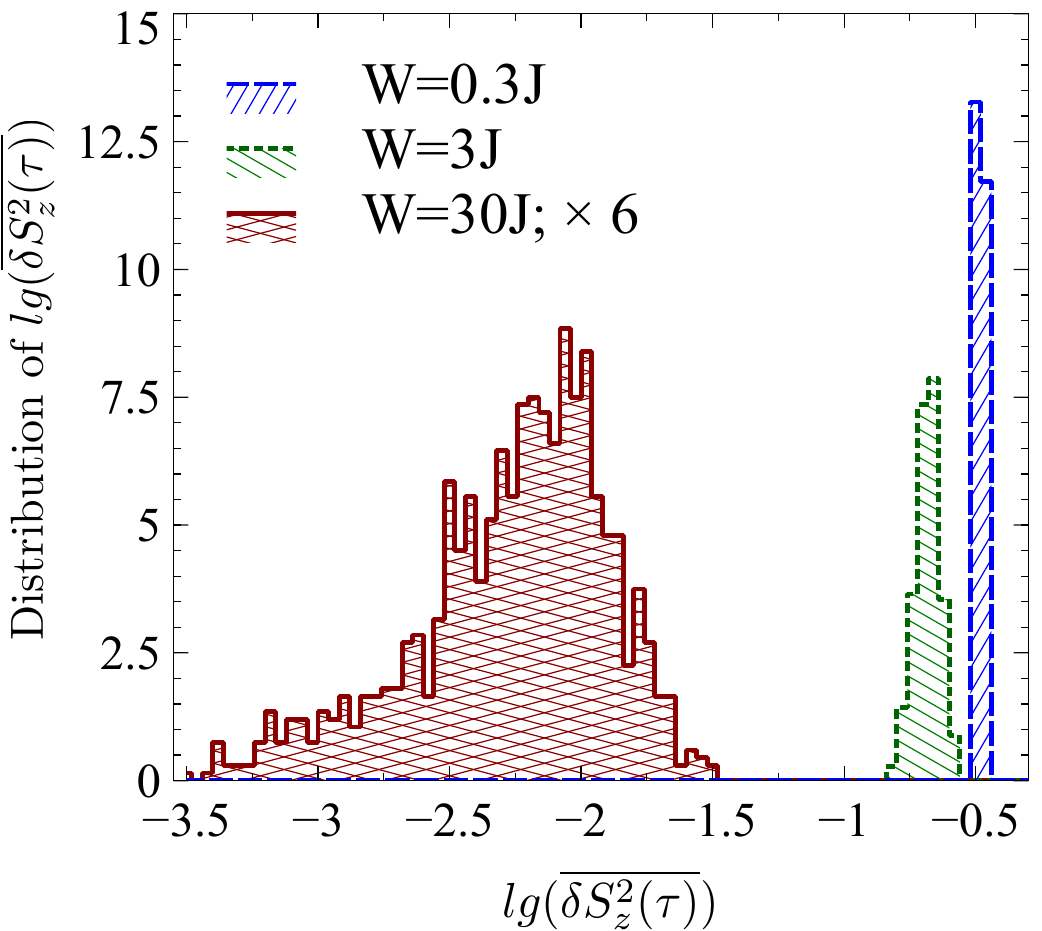}
\par\end{centering}
\caption{\label{fig:Hist_VarSz}
(Color online)
Distribution of $\lg(\overline{\delta S_z^2})$ over $N=10^3$ disorder realizations of the random magnetic field $h_l$ for a system with $L=12$ spins for $W/J=0.3$ (blue long-dashed line) and $W/J=3$ (green short-dashed line), $30$ (red solid line). We scaled the distribution curve for $W/J=30$ by factor six. The averages are performed over all product states of the system.
}
\end{figure}

In Fig.~\ref{fig:Hist_VarSz}, we show the probability distribution of $\lg(\overline{ \delta S_z^2(\tau)})$. The distributions are narrow and the typical value of $\overline{ \delta S_z^2(\tau)}$ is large at weak disorder, whereas the distributions broaden and the typical value of $\overline{ \delta S_z^2(\tau)}$ is small at strong disorder. For the quantum displacement, we showed in the previous section that the distribution of $\lg(\varepsilon)$ is a Pareto distribution. $\lg(\overline{ \delta S_z^2(\tau)})$ distributions for strong disorder have longer tails but not as long as the distributions of quantum displacement $\varepsilon$. However, it is still possible to distinguish between localized and ergodic phases based on $\lg(\overline{\delta S_z^2(\tau)})$ distributions for different disorder strengths even though rare events do not appear and distribution is spread out in a smaller range in the strong disorder.

\begin{figure}[!htbp]
\begin{centering}
\includegraphics[width=\columnwidth]{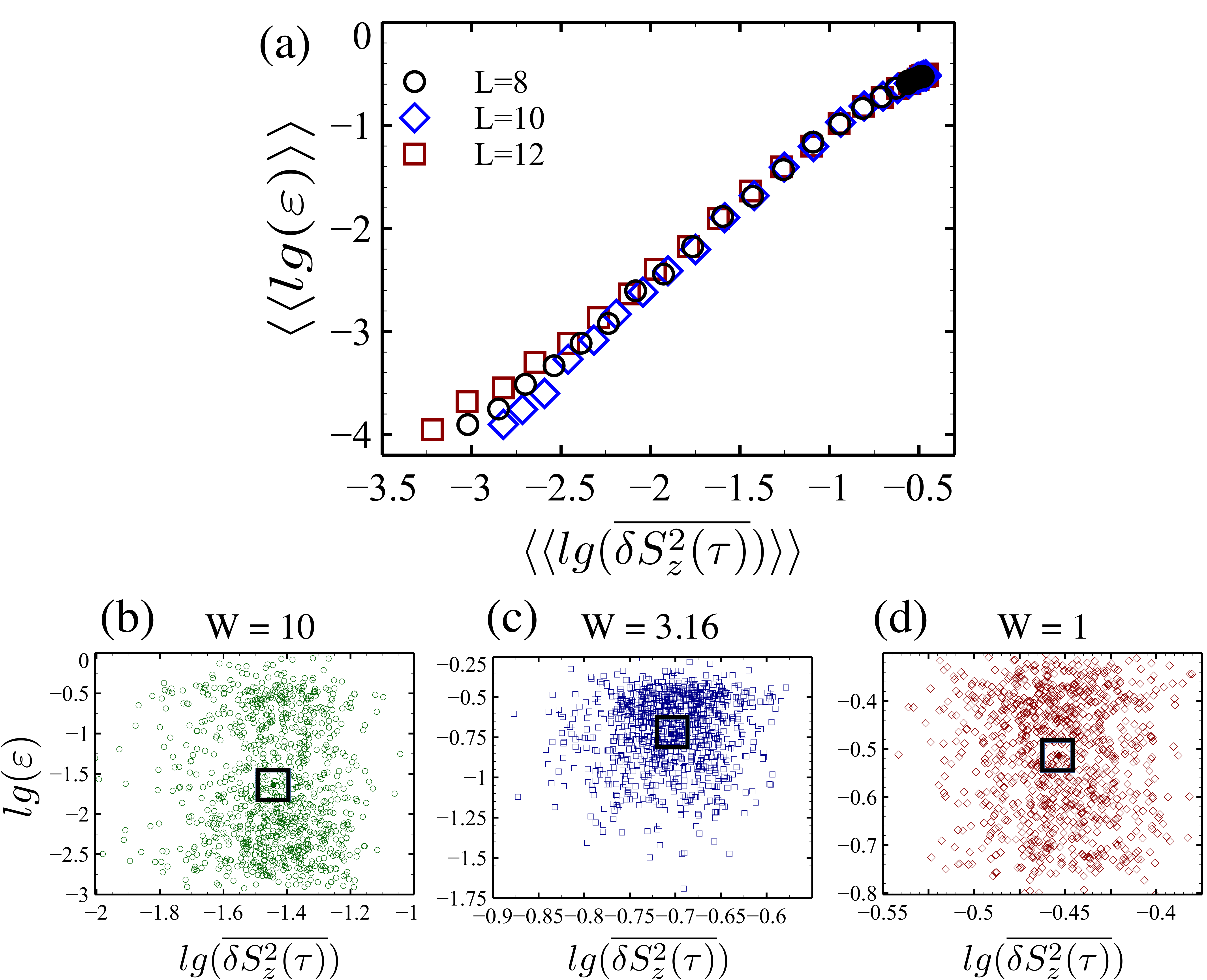}
\par\end{centering}
\caption{\label{fig:ParameterPlot_p_vs_VarSz}
(Color online)
(a) Parameter plot of $\langle\langle \lg(\varepsilon) \rangle\rangle$ and $\langle\langle \lg(\overline{\delta S_z^2}) \rangle\rangle$ for a spin system of size  $L=8$ (circles), $L=10$ (diamonds) and $L=12$ (squares). Data points from Figs.~\ref{fig:avlogT2} and ~\ref{fig:VarSz_vs_W} are used. Time = 1 period. The drive amplitude $f=J/\sqrt{10}$ and $\omega=J$. There is almost a linear dependence between the two quantities. (b, c, d) Scatter plots of the data for the three of the  results for $L = 12$ from (a). $W =$ 1 (red diamond), 3.16 (blue square) and 10 (green circle). Each scatter plot includes $10^3$ unfilled markers each of which corresponds to a single disorder realization. Each average value in plots of (b, c, d) is in a big black square and is shown by a filled marker of same type as the scattered data.}
\end{figure}

We compare the typical values of the  displacement $\varepsilon$ with spin diffusion coefficient $\overline{ \delta S_z^2(\tau)}$. We demonstrate the correlation between $\langle\langle \lg(\varepsilon) \rangle\rangle$ and $\langle\langle \lg(\overline{ \delta S_z^2(\tau)}) \rangle\rangle$ by the parameter plot provided in Fig.~\ref{fig:ParameterPlot_p_vs_VarSz}(a). This behavior of $\langle\langle \lg(\varepsilon) \rangle\rangle$ and $\langle\langle \lg(\overline{ \delta S_z^2(\tau)}) \rangle\rangle$ supports our claim that the total spin measurement can also be used to identify the localization properties of the system. We also provide scatter plots in Figs.~\ref{fig:ParameterPlot_p_vs_VarSz}(b, c, d) for three of the disorder-unaveraged values from Fig.~\ref{fig:ParameterPlot_p_vs_VarSz}(a)  with $W$ = 1 (ergodic regime), 3.16 (critical regime) and 10 (MBL regime). The distributions for both $\lg(\epsilon)$ and $\lg(\delta S_z^2)$ are wide in the localized phase with large disorder strength and the typical values of $\epsilon$ and $\overline{\delta S_z^2}$ are small. For smaller $W$, the distributions get narrower and the typical values are bigger. We deduce from the shape of the clouds in the scatter plots in Figs.~\ref{fig:ParameterPlot_p_vs_VarSz}(b, c, d) that the correlation between $\lg(\epsilon)$ and $\lg(\delta S_z^2)$ are small \cite{Note_CrossCorrelation}. However, as we pointed out, the average values of them are correlated as shown in the parameter plot in Fig.~\ref{fig:ParameterPlot_p_vs_VarSz}(a).

\section{discussion and conclusions}\label{conclusions}

We discussed time dynamics of a quantum Heisenberg spin chain that is subject to a harmonic local drive. We analyzed the overlap  between the states started from the initial states $|\psi_i \rangle$ and evolved under the Heisenberg Hamiltonian with and without drive.  We observed that when averaged over $|\psi_i\rangle$, the quantum displacement after one period $\varepsilon$, given by  Eq.\eqref{eq:QuantumDisplacement}, shows different statistical properties with respect to random field realizations at weak (ergodic) and strong (localized) disorder. As shown in Figs.~\ref{fig:hist} and ~\ref{fig:avlogT2}, in the ergodic regime the distribution of the quantum displacement is narrow and nearly independent from disorder strength, while in the localized regime the distribution has an exponentially 
small average value but a very long power-law tail. The average value of the quantum displacement is independent of system size suggesting that quantum systems with $L\simeq10$ spins would be sufficient to see the distinction between the localized and ergodic regimes using available quantum hardware \cite{schreiber2015observation,Smith2016, Bordia2017, Bernien2017,Neill2018,nguyen2018highcoherence}.

We also studied the variance of the operator for total spin in $z$-direction $\delta S_z^2(t)$, given by Eq.\eqref{eq:Variance_Sz}, for an initial state prepared as a product spin state with total spin projection equal to zero. Thus, $\delta S_z^2(t)$ is a measure of spin accumulation due to the drive and can be used to measure the speed of the thermalization in the ergodic and MBL regimes. Both initialization of this system as a product state of individual spins in $z$-direction and measurement of their net spin projection are basic requirements for quantum hardware and experimental studies of crossover from the ergodic to localized regimes through the spin polarization dynamics is feasible in available systems similar to those described in Refs. \cite{schreiber2015observation,Smith2016, Bordia2017, Bernien2017,Neill2018,nguyen2018highcoherence}.

We calculated the spin accumulation in response to the drive over time $t$, the results are shown in Fig.~\ref{fig:VarSz_vs_time}. In the ergodic regime, the spin accumulation speed is large in the initial periods and total spin gets saturated rapidly. However, in the MBL regime, the spin accumulation is slower in the initial periods and the spins are still drifting in response to the drive in the longer time limit. 
The spin accumulation after one period gives the  spin diffusion coefficient $\delta S_z^2(\tau)$. The behavior of the diffusion coefficient is very similar to the behavior of quantum displacement $\varepsilon$.  As illustrated in Fig.~\ref{fig:VarSz_vs_W}, at weak disorder, diffusion coefficient is large and changes weakly with the disorder strength. However, at strong disorder, the diffusion coefficient decreases linearly with the logarithm of the disorder strength, $lg( W/J)$, and eventually diffusion is broken. The system may show subdiffusive dynamics as recently pointed out in \cite{luitz2019multifractality}. Furthermore, diffusion coefficient does not depend on the system size strongly similar to quantum displacement.

Probability distributions for the diffusion coefficient show different characteristics depending on the disorder strength as can be seen in Fig.~\ref{fig:Hist_VarSz}. At weak disorder, the distribution is narrow and the typical value of the diffusion coefficient is large. At strong disorder, the distribution is wide and have long tail but unlike the distributions for the quantum displacement, the distribution for the diffusion coefficient does not have exponentially long tail and does not exhibit rare events. However, it is still possible to identify the phase of the system based on the diffusion coefficient distributions. The broad distribution of $\delta S_z^2(\tau)$ at strong disorder shows that this parameter cannot be seen as a one-size-fits-all parameter. In other words, there is a different dynamics at strong disorder.

 In Fig.~\ref{fig:ParameterPlot_p_vs_VarSz}, we demonstrated that there is a positive correlation between the quantum displacement and spin accumulation. However, we note that flips of a spin have different effects on the quantum displacement and the  spin accumulation. If a single spin flips, the original and new states, $\ket{\psi}$ and $\ket{\psi'}$ respectively, are orthogonal. 
That makes the displacement $1-|\langle\psi\ket{\psi'}|^2$ between the states equal to 1. However, in the large system size ($L\gg 1$) limit, one spin flip produces a small effect for the total spin $\sim L$ in the  $z$-direction and therefore also for the spin accumulation $\delta S_z^2(t)$. Even though spin flips have smaller effects on the spin accumulation, there is a clear difference between the speed of the thermalization for the two phases as explained above.

Our study was focused on a local harmonic drive with moderate drive frequency ($\omega \simeq J$). For this frequency, we observed that thermalization occurs regardless of whether the system is in the localized or ergodic regimes, which supports the results of \cite{lazarides2014equilibrium, lazarides2015fate, Bordia2017}, and the speed of thermalization is different for the two cases. On the other hand, one could also consider the cases where $\omega$ is much smaller or larger than $J$. In the limit of $\omega<<J$, the time-independent Hamiltonian Eq.~\eqref{eq:transform} will be similar (with difference of $\omega \, S_z$) to the Hamiltonian with dc perturbation considered in \cite{xu2017response}. If the drive frequency is larger than the depth of the local energy minima, different regimes such as prethermal states occur \cite{Else2017, Yao2018}. Most closed many-body systems tend to heat up when they are driven. The situation is different for driven localized systems when many local deep minima appear in the energy spectrum and prevent thermalization. The system is prevented from heating up, which can be understood via quantum mechanics of energy levels. If the drive frequency is large, the system cannot absorb all the energy provided by the drive. Instead, the energy absorption requires many-body excitations 
and slows heating down exponentially \cite{abanin2017effective, abanin2017rigorous}. Under certain nonequilibrium conditions of prethermalization, the systems can exhibit topological phases protected by time-translation symmetry \cite{VonKeyserlingk2016a, Potter2016, Else2016a, Roy2016} and time crystals where time-translation symmetry is spontaneously broken \cite{Khemani2016, VonKeyserlingk2016_timecrystal, Else2016, Else2017, Yao2017, Zhang2017, Choi2017, Yao2018, machado2019prethermal}. Exploring statistics of the system responses at high frequency periodic drive was not addressed here and is the topic of a separate study. \newline

\begin{acknowledgments}
We thank D. Basko, M. Dykman, D. Huse, L. Ioffe, I. Martin, R. Nandkishore and V. Oganesyan for fruitful discussions. This work was supported by the U.S. Department of Energy, Office of Science, Office of Basic Energy Sciences, under Award Number DE-SC0019449. The simulations were performed using the computing resources of the UW-Madison Center For High Throughput Computing (CHTC) and resources provided by the Open Science Grid \cite{pordes2007open,sfiligoi2009pilot}, which is supported by the National Science Foundation award 1148698 and the U.S. Department of Energy's Office of Science. Numerical simulations were performed using QuTiP \cite{johansson2013qutip}.

\end{acknowledgments}


\begin{thebibliography}{77}
\bibitem[{Not({\natexlab{a}})}]{NoteFloquet}
\bibinfo{note}{We note that the Floquet Hamiltonian is equal to $\tilde H$ for
  periodic drive given in Eq.(3).}
\bibitem[{Not({\natexlab{c}})}]{Note_Integrability}
\bibinfo{note}{We also compared the spin accumulation for $W$ = 0 (zero
  disorder) with the ergodic ($W$ = 1.25) and MBL ($W$ = 5) cases. We observed
  that when there is no disorder, the behavior of variance is quite different
  from the evolution in the ergodic regime, so $W = 0$ case cannot simply be
  studied by choosing $W \longrightarrow 0$ and deserves a separate analysis.}
\bibitem[{Not({\natexlab{d}})}]{Note_CrossCorrelation}
\bibinfo{note}{We used cross-correlation to calculate the correlation between
  $\lg(\epsilon)$ and $\lg (\overline{\delta S_z^2})$.}
\end{thebibliography}

\end{document}